\documentclass[12pt]{iopart}
\usepackage{epsfig}
\usepackage{graphicx}
\usepackage{lscape}
\usepackage{longtable}

\usepackage{iopams}  
\begin{document}

\title[V488 Per]{Warm Dusty Debris Disks and Distant Companion Stars:  V488 Per and 2M1337}

\author{B. Zuckerman$^1$}

\address{$^1$Department of Physics and Astronomy, University of California, Los Angeles, CA 90095, USA}
\eads{\mailto{ben@astro.ucla.edu}}
\begin{abstract}
A possible connection between the presence of large quantities of warm (T$\geq$200 K) circumstellar dust at youthful stars and the existence of wide-separation companion stars  has been noted in the literature.  Here we point out the existence of a distant companion star to 
V488 Per, a K-type member of the $\alpha$ Persei cluster with the largest known fractional excess infrared luminosity ($\sim$16\%) of any main sequence star.  We also report the presence of a distant companion to the previously recognized warm dust star 2M1337.  With these discoveries the existence of a cause and effect relationship between a distant companion and large quantities of warm dust in orbit around youthful stars now seems compelling.  
\end{abstract}

\section{INTRODUCTION}

Dusty debris disks around main sequence stars were discovered by the Infrared Astronomical Satellite (IRAS; Aumann et al 1984).   It was soon understood that the observed dust particles are a consequence of destructive collisions of larger (unseen) objects and are not a leftover remnant of an early protoplanetary phase of stellar evolution.  The disks are typically dominated by cool dust (T $<$100 K) that emits at far-infrared wavelengths although some much warmer dust is often also seen (e.g., Morales et al 2011).   Much less common are debris disks where most of the infrared emission appears in the mid-infrared and is emitted by warm dust particles with temperatures $\geq$200 K (e.g., Melis et al 2010).   One can characterize the infrared dust luminosity as a fraction of the bolometric luminosity (L$_{bol}$) of the central star.  For warm dust stars such as those discussed by Melis et al, of order half or more of the fractional dust luminosity L$_{IR}$/L$_{bol}$ emerges at mid-IR wavelengths.   
In the present paper to be defined as a warm dust star the fractional excess IR luminosity must be of order 1\% or more; the motivation for this limit is given in Section 2.   

It has been recognized for some time that warm dust stars are often found in multiple star systems (e.g., Kastner et al 2012).   Here we report the likely existence of one or two distant companions  to V488 Per a warm dust K3-type member of the $\alpha$ Persei  cluster (Zuckerman et al 2012).  We also report the existence of a distant companion to warm dust M3.5-type star 2MASS J13373839-4736297 (called 2M1337 by Schneider et al 2012b).  As discussed in Section 3.3, the total known sample of warm dust stars with distant companions is now sufficiently large (Tables 1 and 2) that there must be a causal connection between these two phenomena.

\section{SAMPLE SELECTION}

A systematic search for youthful warm dust stars in the IRAS catalog was
carried out by C. Melis and collaborators (Melis 2009).  This program yielded
the warm dust stars HD 23514 (Rhee et al 2008), HD 15407 (Melis et al
2010), and HD 131488 (Melis et al 2013).  Various other warm dust stars were identified, but these turned out
to be first ascent giant stars (Melis et al 2009; Melis 2009).  Other than
HD 23514, HD 15407 and BD+20 307 (Zuckerman et al 2008), it appears that no main sequence star of 
age $>$35 Myr and spectral type F and later in the IRAS catalog has 
L$_{IR}$/L$_{bol}$ of order 1\% or greater.

Other surveys that would have revealed warm dust stars have been carried out with the Spitzer Space Telescope and with the Wide-field Infrared Survey Explorer (WISE).   Such searches have revealed three additional warm dust stars -- P1121, ID8, and V488 Per -- with ages $\geq$35 Myr (see Table 1). 


With only one exception -- BD+20 307, which is composed of two solar-type stars in a close orbit (Weinberger 2008) -- no known warm dust star is older than about 100 Myr.   Since there are far more stars in the solar vicinity with ages $>$100 Myr than $<$100 Myr, Melis et al (2010) argued that the warm dust phenomenon is associated with the last stages of rocky planet formation in the terrestrial planet zone.  Specifically, if one assumes a constant star formation rate in the Milky Way during the past two Gyr, then there are 20 times more solar-like stars with ages between about 100 Myr and two Gyr than there are between 35 and 120 Myr; but there are 5 times as many warm dust stars in the latter age group than in the former.   

Spitzer and WISE were sufficiently sensitive to detect warm dust with infrared luminosity more than an order of magnitude smaller than 1\% of L$_{bol}$.   Such luminosities can be radiated by dust produced in the collision of two modest size asteroids.  Even at dust luminosities as small as 10$^{-4}$ L$_{bol}$, for stars older than 100 Myr, very few warm dust stars have been reported (for example, HD 69830 and HD 169666; Beichman et al 2005; Moor et al 2009; Olofsson et al 2012).  Rhee et al (2008) and Melis et al (2010) argue that to produce L$_{IR}$/L$_{bol}$ of order 1\% or greater, individual colliding bodies would have masses of order that of Earth's moon, i.e. characteristic of planetary embryos.  

In Table 1 we restrict the sample to stars with ages between about 30 and 120 Myr, dust luminosities of $\geq$1\%, and dust temperatures of order 200 K or greater.  In this way we are considering the most collisionally active debris disks in the terrestrial planet zone in the latter stages of rocky planet formation.  
Neither A-type nor M-type warm dust stars appear in Table 1, i.e., none are known with ages 30 Myr or greater that satisfy our dust temperature and luminosity constraint.   This is consistent with previous studies.  For example, Deacon et al (2013) review the literature on debris disks (both warm and cool) around late-type stars and conclude "in summary, candidate debris disks around late-K and M dwarfs are rare after 20 Myr." The reason for this rarity is not yet understood.  

For ages $<$30 Myr warm dust stars appear with spectral types ranging from A- to M-type.   We have excluded solar type star HD 166191 from Table 2 because it may be substantially younger than 10 Myr (Kennedy et al 2014 and references therein).  
We choose to limit our study to stars of approximately solar mass and lower and thus do not consider A-type stars with dominant warm dust emission.  Four A-type stars with ages $\sim$10 Myr -- HD 131488, 172555, HD 121191, and EF Chamaeleontis -- are considered by Melis et al (2013).  (HD 172555 and EF Cha have fractional IR luminosities well below our approximate 1\% cutoff.)  

Our analysis 
also excludes systems that contain both a wide companion and a spectroscopic binary with warm circumbinary dust; this pertains specifically to two warm dust stars in the TW Hya Association (TWA) and one in the $\beta$ Pictoris moving group.  This exclusion is motivated by uncertainty in the role the close binarity might play in generation of massive quantities of warm dust.  Hen3-600 is a triple system in the TWA where
the circumbinary IR emission is dominated by 200 K dust (e.g., Figure 2 in Zuckerman 2001).  Similarly, TWA member HD 98800 is a quadruple system containing two sets of spectroscopic binaries.  Its dominant dust emission is at 160 K (Figure 1 in Zuckerman \& Becklin 1993), and the warm dust orbits one of the spectroscopic binaries.   V4046 Sgr is a K-type, warm dust, spectroscopic binary in the $\beta$ Pictoris moving group that likely has an M-type spectroscopic binary companion at a projected separation of $\sim$12,300 AU (Kastner et al 2011).

\section{DISCUSSION}

First we consider evidence for companions to V488 Per which is the dustiest main sequence star currently known (L$_{IR}$/L$_{bol}$ $\sim$16\%; Zuckerman et al 2012).   Then we describe a  companion to previously known warm dust star 2M1337.  Finally we consider the ensemble of warm dust stars and argue the case for their preferential existence in wide multiple star systems.    

\subsection{V488 Per Is a Member of a Wide Multiple Star System}

Randich et al (1996) carried out an extensive X-ray survey of the $\alpha$ Persei open cluster.  About 160 X-ray sources were detected of which 89 could be identified with known objects.  However, no known optical counterparts were found for 73 sources.   Prosser \& Randich (1998; hereafter PR1998) used optical photometry and spectroscopy to determine which of these 73 so-called "APX" X-ray sources are associated with members of $\alpha$ Per.   The result was $\sim$40 new candidate cluster members in the late-G to M dwarf range.

A figure in the Appendix of PR1998 presents finding charts of the vicinity of all 73
X-ray fields examined in their search for $\alpha$ Per member counterparts.  The charts are 3 x 3 arcmin on a side centered on the X-ray source position.   PR1998 searched for optical counterparts within 30" of the X-ray source positions.

Field $\#$43 is the only one in which a previously known $\alpha$ Per member (that is not one of the 73 X-ray sources, see discussion below) is seen.   In field $\#$43 V488 Per (= AP 70) is seen 71" from APX43A and 69" from APX43B both of which are classified as definite cluster members in PR1998.    In the following we describe why at least one of these APX stars is likely to be a physical companion of V488 Per.   

The $\alpha$ Persei cluster contains of order 200 K- through B-type known members (Prosser 1992; Zuckerman et al 2012) within a radius of 3 degrees of the cluster center at 03h26m, +49d07'.  Given that V488 Per is located about 1/2 degree from the cluster center, we adopt a surface density of 20 K- through B-type stars per square degree.  Then the probability that two K-type members (V488 Per and APX43A) not physically bound to each other lie within 71" of each other in the plane of the sky is $\sim$2.5\%.   If the $\alpha$ Per cluster luminosity function is similar to that of field stars (Prosser 1992; Lodieu et al 2012), then the probability that a K-type member (V488 Per) and a mid-M type member (APX 43B) are within 69" of each other is $\sim$5\%.

Of the 40 or so candidate new cluster members identified in PR1998, none but field $\#$43 that contains APX43A and APX43B also contains a previously known cluster member (see the two paragraphs that follow).  Thus one can safely say that the probability of a chance sky plane alignment of two unrelated $\alpha$ Persei cluster members within a 3' x 3' field of view of the Prosser \& Randich search is small, as is the frequency of cluster binary stars with separations of order an arc minute.   This result is consistent with the probabilities estimated in the previous paragraph.

While a cursory inspection of Table 2 in PR1998 might lead one to believe that fields other than $\#$43 contain multiple members of the $\alpha$ Per cluster, such is not the case.  
In field $\#$1 previously known $\alpha$ Per member AP 101 is indicated near star APX3D which PR1998 classify as a probable (Y?) member of $\alpha$ Per.   However, AP 101 is misplotted (it is far from APX3D) and APX3D is definitely not a member of the $\alpha$ Per cluster based on its proper motion given in both the PPMXL and UCAC4 catalogs.  Rather, based on color information given in PR1998 and in catalogs in Vizier, APX3D is an M-type star much nearer to Earth than is the $\alpha$ Per cluster.  The M-type nature of APX3D is confirmed by a recent spectrum obtained with the HIRES spectrometer on the Keck telescope (L. Vican et al, in preparation).  

In Table 2 in PR1998, when an APX $\#$ is replaced by an HE $\#$, then the X-ray source is associated with a previously known cluster member and no new member has been identified by PR1998.  Thus, for example, APX21 is actually previously known HE 606 and HE 604 and 600 are not members.  Similarly, APX64 is actually previously known member HE 955.  Finally, APX55 is identified in PR1998 as a possible cluster member near HE 879.  However, studies subsequent to 1992 indicate that HE 879 is not a cluster member (e.g., Tables 1 and 2 in Zuckerman et al 2012).

Thus, in a sample of 40 probable cluster members identified in the PR1998 survey, only APX43A and APX43B are located in the plane of the sky near a previously known cluster member (V488 Per).  Therefore, if either or both of APX43A and APX43B are cluster members, then they are very unlikely to be members far from V488 Per that just happen to lie along similar lines of sight to Earth.

APX43A and APX43B are, respectively, 2MASS J03281724+4840577 and J03281658+4840542 (Skrutskie et al 2006).   Based on color information given in PR1998 and in Vizier, we estimate their spectral types to be K4 and M4.5.  Entries in the UCAC4 and UKIDSS DR9 catalogs indicate that APX43A has, within the errors, the same proper motion as V488 Per (Table 3) while its colors and brightness in the 2MASS, UCAC4, UKIDSS DR9, and Initial Gaia Source List ("IGSL", Smart 2013) catalogs are consistent with those of V488 Per.   Thus APX43A is likely to be a cluster member and a companion of V488 Per.   

Proper motions for APX43B are given in UKIDSS DR9 and in the IGSL and both agree with the proper motion of V488 Per (Table 3).  However the error bars on the IGSL values are so large as to render this agreement of little value.   At K-band APX43B is one magnitude fainter than V488 Per and APX43A; this probably is consistent with a common distance (e.g., Figure 5 in Rodriguez et al 2013).  Lodieu et al (2012) give APX43B a 47\% chance to be a member of the $\alpha$ Per cluster.  Thus, APX43B may be a member of the $\alpha$ Per cluster and, if it is, then as noted earlier in this section it is likely a physically bound companion of V488 Per.

In summary, V488 Per likely has one distant companion, APX43A, and may have another one too (APX43B).  
Since the sky plane separation between APX43A and 43B is $\sim$7" and between V488 Per and the two APX stars $\sim$70", there is no obvious problem with dynamical stability should the system be a triple.

\subsection{A Companion to 2MASS J13373839-4736297}

Schneider et al (2012b) present an SED for warm dust star 2MASS J13373839-4736297 which they dub 2M1337.  We now note the existence of a star 2MASS J13373825-4736397 of similar apparent brightness located $\sim$10" south of 2M1337.   Based on inspection of the PPMXL, SPM4.0, UCAC4, 2MASS and DENIS catalogs, it is obvious that the two stars comprise a binary system.   We therefore rename 2M1337 as 2M1337A and its companion 2M1337B, while retaining the name 2M1337 to mean the entire binary system.  

From consideration of the ALLWISE magnitudes for 2M1337B and Table 6 in Pecaut \& Mamajek (2013), one might be led to conclude that there is excess infrared emission in all four WISE bandpasses.  However, inspection of the magnitudes of 2M1337B in the WISE all-sky survey indicate that all of the apparent IR excess in ALLWISE is actually due to  blending in of some emission from 2M1337A (R. Cutri 2014, private communication).

\subsection{Are stars with warm dusty debris disks preferentially located in wide-orbit multiple star systems?}

Tables 1 and 2 list a majority of warm dust stars presently known but, as noted in Section 2, some warm dust stars have been purposely excluded.  
In the future, wide-orbit companions to some of the apparently single warm dust stars listed in these tables may be uncovered.  Indeed the wide-orbit companions  to V488 Per, HD 23514, TW Hya, and 2M1337A were identified only years after their dusty nature was recognized.  Thus the percentages of warm dust systems in Tables 1 and 2 deemed to be in a wide-orbit binary should be regarded as lower limits.

Given the large percentage of wide-orbit binary stars in Tables 1 and 2, a question of interest is whether warm dusty debris disks are preferentially associated with this type of binary system.  Some insight may be provided from recent reviews of stars in multiple systems (Duchene \& Kraus 2013) and in young multiple systems (Reipurth et al 2014).   We first consider stars of roughly solar mass (third column of Table 1).  Papers cited in both reviews indicate a single star frequency in the field of 55\% plus/minus a few percent.  
Figure 2 in Duchene \& Kraus (2013; hereafter DK2013) presents number of binaries vs. semi-major axis as a function of spectral type.   Binary stars with semi-major axes between 300 and 10000 AU -- the range covered by the stars in Table 1 -- comprise about 25\% of the total of those with G-type primaries.  If these were the only considerations then about one system in 8 would be a multiple with semi-major axis in the range 300-10000 AU.   However, as seen in Figure 5 and Table 2 of DK2013, for solar-type stars with ages 30-100 Myr such as those in Table 1 of the present paper, with considerable uncertainty, wide binaries might be twice as common relative to single stars as they are among older field stars.   Thus, for 30-100 Myr old solar-type stars, one anticipates that about one in four (25\%) would be a binary with semi-major axis between 300 and 10000 AU.  But 60\% (3 in 5) of the warm dust systems in Table 1 are located in such a binary system; the probability for this large a fraction is only 9\%. This comparison suggests that solar-type warm dust systems are found preferentially in wide-orbit multiple star systems but, because of small number statistics, the evidence is not compelling.

Far more compelling evidence for a causal relationship between warm dusty disks and a wide-orbit companion comes from Table 2.  About 1/4 of M-type field stars are to be found in multiple systems (DK2013, Table 1 and Figure 1).  From Figure 2 in DK2013, $<$1\% of M-type binaries have semimajor axes $\geq$1000 AU.  From Figure 5 and Table 2 of DK 2013 we estimate that, among 10-20 Myr old M-type stars, wide-binary systems might be about 4 times as frequent relative to single stars as they are among much older field stars.   Combination of these three factors suggests that, 
if the warm dust phenomenon were randomly associated with single stars and with binary stars of various separations, then perhaps one percent or so of warm dust 10-20 Myr old M-type stars should be found in multiple systems with semi-major axes $\geq$1000 AU.  Yet among the 10 K7-M6 warm dust stars in Table 2, at least 6 are in binary systems with semi-major axes $>$1000 AU, the probability of which is 2 x 10$^{-4}$.

The stars in Table 1 are apt to be examples of dusty debris disks that contain relatively little gas.  The basis for this statement is their (relatively old) age and the absence of any evidence for the presence of orbiting or accreting gas.   The same cannot be said for many of the much younger stars in Table 2.   The column in Table 2 headed "acc.?" indicates whether a star is currently accreting material from a surrounding disk (y = yes; n = no).  In addition to references listed in the table, evidence regarding accretion at TWA 31 can be found in Shkolnik et al (2011), at LDS 5606A and B in Zuckerman et al (2014), and at 2M1337A in Rodriguez et al (2011) and Schneider et al (2012b).  The presence of accreting gas suggests a possibly substantial reservoir of gaseous material, material that may be left over from a protoplanetary phase of evolution.  Thus, some of the stars listed in Table 1 may not be pure debris disks of the sort considered in the first paragraph of the Introduction.   But a correlation between large quantities of warm dust and a distant companion star remains.

\section{CONCLUSIONS}

Excess infrared emission above the photospheres of numerous main sequence stars indicates that  
dusty debris disks are common.  But for only a small percentage does the dominant excess emission appear at mid-IR rather than far-IR wavelengths.  Dominant mid-IR emission requires warm dust particles (T$\geq$200 K) and indicates orbital semi-major axes between about one AU and as little as a few hundredths of an AU, depending on the specific dust temperature and the luminosity of the central star.

In the present paper we consider stars with excess infrared emission that is dominated by warm dust particles and with spectral type M through mid-F and ages 10-100 Myr.
We find that two such warm dust stars have distant companions not previously recognized.  By adding these examples to those previously known, one can show that the existence of the warm dust phenomenon is strongly correlated with the presence of a distant companion star (typical semi-major axis $\geq$1000 AU).  Thus the ratio of star-star separation to warm dust particle orbital semi-major axes may often be as large as 10$^4$.  Questions to be answered include:  how exactly does the gravity of such a distant star so effectively stir up the dynamics of dust particles and/or larger (unseen) objects located so close to a companion star and what do such dynamics imply for rocky planet formation in the terrestrial planet zone?  A potential complication is the possibility that some of the dust particles in the younger disks are not second generation debris from collisions of large objects but rather might be material left over from a gaseous protoplanetary phase of stellar evolution.

$\smallskip$

We thank Dr. Carl Melis for very helpful comments, Dr. Adam Schneider for very helpful assistance, and the referee for suggestions that improved the paper.  This research was supported by NASA grants to UCLA.

\section*{REFERENCES}
\begin{harvard}

\item[Aumann, H., Beichman, C., Gillett, F. et al 1984, ApJ 278, L23]
\item[Beichman, C., Bryden, G., Gautier, T. et al 2005, ApJ 626, 1061]
\item[Deacon, N., Schlieder, J., Olofsson, J., Johnston, K. \& Henning, Th. 2013, MNRAS 434, 1117]
\item[Duchene, G. \& Kraus, A. 2013, ARA\&A 51, 269]
\item[Gorlova, N., Padgett, D., Rieke, G. et al. 2004, ApJS 154, 448]
\item[Gorlova, N., Balog, Z., Rieke, G. et al 2007, ApJ 670, 516]
\item[Kastner, J., Sacco, G., Montez, R et al 2011, ApJ 740, L17].
\item[Kastner, J. Thompson, E., Montez, R. et al 2012, ApJ 747, L23]
\item[Kennedy, G. Murphy, S., Lisse, C. et al 2014, MNRAS 438, 3299]
\item[Lodieu, N., Deacon, N., Hambly, N. \& Boudreault, S. 2012, MNRAS 426, 3403]
\item[Looper, D., Mohanty, S., Bochanski, J. et al 2010a, ApJ 714, 45]
\item[Looper, D., Bochanski, J., Burgasser, A. et al 2010b, AJ 140, 1486]
\item[Melis, C. 2009, Ph.D. thesis, UCLA]
\item[Melis, C., Zuckerman, B., Rhee, J. \& Song I. 2010 ApJ 717, L57]
\item[Melis, C., Zuckerman, B., Rhee, J. et al 2012, Nature 487, 74]
\item[Melis, C., Zuckerman, B., Rhee, J. et al 2013, ApJ 778, 12]
\item[Melis, C., Zuckerman, B., Song, I., Rhee, J. \& Metchev, S. 2009, ApJ 696, 1964]
\item[Moor, A., Apai, D., Pascucci, I. et al 2009, ApJ 700, L25]
\item[Morales, F., Rieke, G. Werner, M. et al. 2011, ApJ, 730, L29]
\item[Olofsson, J., Henning, Th., Nielbock, M. et al. 2013, A\&A 551, A134]
\item[Olofsson, J., Juhasz, A., Henning, Th. et al 2012, 542, A90]
\item[Pecaut, M. \& Mamajek, E. 2013, ApJS, 208, 9]
\item[Prosser, C. 1992, AJ 103, 488]
\item[Prosser, P. \& Randich, S. 1998, AN 319, 201 (PR1998)]
\item[Randich, S., Schmitt, J., Prosser, C. \& Stauffer, J. 1996, A\&A 305, 785]
\item[Reipurth, B., Clarke, C., Boss, A. et al 2014, arXiv:1403.1907]
\item[Rhee, J., Song, I. \& Zuckerman,, B. 2008, ApJ 675, 777]
\item[Rodriguez, D., Bessell, M., Zuckerman, B. \& Kastner, J. 2011, ApJ 727, 62]
\item[Rodriguez, D., Marois, C., Zuckerman, B., Macintosh, B. \& Melis, C. 2012, ApJ 748, 30]
\item[Rodriguez, D., Zuckerman, B., Faherty, J. \& Vican, L. 2014, A\&A 567, 20]
\item[Rodriguez, D., Zuckerman, B., Kastner, J. et al 2013, ApJ 774, 101]
\item[Schneider, A., Song, I. \& Melis, C. 2012a, ApJ 754, 39]
\item[Schneider, A., Song, I., Melis, C., Zuckerman, B. \& Bessell, M. 2012b, ApJ 757, 163]
\item[Shkolnik, E., Liu, M., Reid, I. N., Dupuy, T. \& Weinberger, A. 2011, ApJ 727, 6]
\item[Skrutskie, M., Cutri, R., Stiening, R. et al 2006, AJ 131, 1163]
\item[Smart, R. 2013, VizieR On-line Data Catalog: I/324]
\item[Teixeira, R., Ducourant, C., Chauvin, G. et al 2008, A\&A 489, 825]
\item[Weinberger, A. 2008, ApJ 679, L41]
\item[Zuckerman, B. 2001, ARA\&A 39, 549]
\item[Zuckerman, B. \& Becklin, E.E. 1993, ApJ 406, L25]
\item[Zuckerman, B., Fekel, F., Williamson, M., Henry, G. \& Muno, M. 2008, ApJ 688, 1345]
\item[Zuckerman, B., Melis, C., Rhee, J., Schneider, A. \& Song, I. 2012, ApJ 752, 58]
\item[Zuckerman, B., Vican, L. \& Rodriguez, D. 2014, ApJ 788, 102]

\end{harvard}

\clearpage

\begin{table}
\caption{Solar-type Stars of Age 35 - 100 Myr With Dominant Warm Dust Emission}
\begin{tabular}{@{}lccccccc}
\br
name & group  & spec & Dist. &  comp. & separation & age  &  ref. \\
 & & type & (pc) & type &  (AU) & (Myr) &   \\
\mr
V488 Per & $\alpha$ Per & K3 & 182 & K4/M4.5 & 12700  &   90  & this paper \\
HD 15407 & AB Dor &  F5V & 55 &  K2V & 1160 & 100 & Melis 2010 \\
HD 23514  & Pleiades  &  F6  & 136 & M7 & 360 & 120 & Rodriguez 2012 \\
P1121 &  M47 & F9 IV/V & 450 &  &   &  100 &   Gorlova 2004\\
ID8  &  NGC 2547 & G6V  & 430  &    &   &  35 &  Gorlova 2007\\
\br
\end{tabular}
\end{table}

\begin{table}
\caption{Stars of Age 10-20  Myr With Dominant Warm Dust Emission}
\begin{tabular}{@{}lcccccccc}
\br
name & group  & spec & Dist. &  comp. & separation & age  & acc.?  & ref. \\
 & & type & (pc) & type &  (AU) & (Myr) &   \\
\mr
LDS 5606A  &  $\beta$ Pic  &  M5  & 65 & M5 & 1700 & 20 & y  & Rodriguez 2014 \\
LDS 5606B  &  $\beta$ Pic  &  M5  & 65 & M5 & 1700 & 20 &  y?  &Rodriguez 2014 \\
TWA 34  &  TWA  &  M5  &  50  &   &  &  10 & n? &  Schneider 2012b \\
TW Hya &  TWA  &  K7  &    55 &  M8.5  &  41000 & 10  &  y  & Teixeira 2008 \\
TWA 30A  &  TWA  & M5  &  42  & M4  & 3400  &  10  &  y  & Looper 2010b \\
TWA 30B  &  TWA  & M4 &  42 & M5  &  3400 & 10  &  y  &  Looper 2010b\\
TWA 33  &  TWA  &  M5  &  42  & &  & 10  &  n?  & Schneider 2012b\\
TWA 31 &  TWA & M4  &  110  &    &  & 10 &  y  &Schneider 2012a \\
TWA 32  & TWA  &  M6 & 53 & & &  10  & n  &   Schneider 2012a \\
TYC 8241* & LCC & K2 &  140 &  &  & 12  & n  &  Melis 2012\\
HD 113766 & LCC & F3 & 120 &  F5 & 160 & 12 &  n   & Olofsson 2013 \\
2M1337A &  LCC &  M3.5  &  120 &  M3.5  & 1200 & 12 &  n?  &  this paper \\
\br
\end{tabular}
Notes - *TYC 8241 = Tycho 8241-2652-1; LCC = Lower Centaurus-Crux  \\
\\
  \\
  \\
 \\
  \\
  \\
  \\
  \\
\end{table}

\clearpage

\begin{table}
\caption{Proper Motions of V488 Per, APX43A, and APX43B}
\begin{tabular}{@{}lccc}
\br
star & UCAC4 &  UKIDSS DR9   &  average \\
     & \ \ $\mu_{\alpha}$ \ \ \ \ \ $\mu_{\delta}$  & \ \  $\mu_{\alpha}$ \ \ \ \ \  $\mu_{\delta}$   & \ \  $\mu_{\alpha}$ \ \ \ \ \ $\mu_{\delta}$  \\
\mr
V488 Per &  \ \ \ 20.1 \ \  -25.0 & \ \ \ 25.7  \ \ -23.2 &  \ \ \ 22.9  \ \  -24.1 \\
APX43A  &  \ \ \ 26.0  \ \  -19.0  &  \ \ \ 25.0 \ \  -19.8  & \ \ \ 25.5  \ \ -19.4 \\
APX43B  &    &  \ \ \ 22.4  \ \ -19.4 & \ \ \  22.4  \ \  -19.4 \\
\br
\end{tabular} 
 \end{table}
Notes - The listed proper motions are in mas/yr and the errors quoted in the two catalogs are all close to $\pm$2.2 mas/yr in both R.A. and Decl.  The 4.7 mas/yr difference between the average declination proper motion of V488 Per and those of the two APX43 stars is obviously within the measurement errors since for V488 Per itself the difference in the R.A. proper motion measured by UCAC4 and UKIDSS is 5.6 mas/yr.

\end{document}